\documentclass[aps,prl,showpacs,superscriptaddress]{revtex4}
\usepackage{graphicx}
\usepackage{bm}
\usepackage{color}

\begin{document}

\title{Dominance of Radiation Pressure in Ion Acceleration with Linearly Polarized Pulses at Intensities of $10^{21}\textrm{W}\textrm{cm}^{-2}$}
\author{B. Qiao}
\affiliation{School of Mathematics and Physics, Queen's University Belfast, Belfast BT7 1NN, UK}
\affiliation{Center for Energy Research, University of California San Diego,  La Jolla, CA 92093 USA}
\author{S.Kar}
\author{M. Geissler}
\affiliation{School of Mathematics and Physics, Queen's University Belfast, Belfast BT7 1NN, UK}
\author{P. Gibbon}
\affiliation{J\"ulich Supercomputing Center, Forschungzentrum J\"ulich GmbH,  
D-52425, J\"ulich, Germany}
\author{M. Zepf}
\author{M. Borghesi}
\affiliation{School of Mathematics and Physics, Queen's University Belfast, Belfast BT7 1NN, UK}

\begin{abstract}

A novel regime is proposed where, employing linearly polarized laser pulses at intensities $10^{21}\textrm{Wcm}^{-2}$ as two order of magnitude lower than earlier predicted [T. Esirkepov et al., Phys. Rev. Lett. 92, 175003 (2004)], ions are dominantly accelerated from ultrathin foils by the radiation pressure, and have monoenergetic spectra. In the regime, ions accelerated from the hole-boring process quickly catch up with the ions accelerated by target normal sheath acceleration (TNSA), and they then join in a single bunch, undergoing a hybrid Light-Sail/TNSA acceleration. Under an appropriate coupling condition between foil thickness, laser intensity and pulse duration, laser radiation pressure can be dominant in this hybrid acceleration. Two-dimensional PIC simulations show that $1.26\textrm{GeV}$ quasimonoenergetic $\textrm{C}^{6+}$ beams are obtained by linearly polarized laser pulses at intensities of $10^{21}\textrm{Wcm}^{-2}$.

\end{abstract}

\pacs{52.38.Kd, 52.50.Jm, 29.25.-t, 52.65.Rr}
\maketitle

Laser-driven ion acceleration from solid foils has attracted great interest due to its many prospective applications \cite{Macro, Malka} including isotope production, tumor therapy, ultrafast radiography, and inertial confinement fusion. Several of these applications require a high-energy ion beam with a large particle number and a monoenergetic spectrum. Currently two main acceleration mechanisms have been identified and widely investigated: target normal sheath acceleration (TNSA) \cite{previous,Nature(2006)} and radiation pressure acceleration (RPA) \cite{Macchi(2005),Robinson(2008),Bin(2009),Macchi(2009),Bin(2010)}. In TNSA, the acceleration of ions is due to the strong electrostatic sheath field at the rear of the foil, created by hot electrons generated from the foil front side via the oscillating $\mathbf{j}\times\mathbf{B}$ heating of linear polarized (LP) lasers. The ion beams produced by TNSA are typically characterized by low particle density, large divergence and broad energy spread. RPA in principle is a very efficient scheme for obtaining high-quality monoenergetic ion beams, via the so-called "hole-boring" (HB) \cite{Robinson(2008)} and "light-sail" (LS)\cite{Macchi(2009),Bin(2009)} stages. It is generally accepted that an effective implementation of the RPA scheme at realistic irradiances is facilitated by the use of circularly polarized (CP) laser pulses \cite{Robinson(2008),Macchi(2009),Bin(2010)}, as this will provide a constant ponderomotive drive. This non-oscillating ponderomotive force applies a steady pressure to the foil front surface and piles up electrons into a compressed layer, inducing an intense charge separation field that accelerates the ions. This process, known as HB, is repeated till the laser pulse punches through the foil. At later times the compressed electron and ion layers combine together to form a quasineutral plasma slab undergoing LS RPA \cite{Macchi(2009)}. 

By contrast, for LP lasers the oscillating $\mathbf{j}\times\mathbf{B}$ term causes heating and recirculation of hot electrons through the target, resulting in a large sheath field at the rear side of the foil. The ions at the rear are accelerated by TNSA and the foil undergoes rapid decompression, breaking the equilibrium condition \cite{Macchi(2009)} required for LS RPA. A RPA dominated regime ("laser-piston") has been identified by Esirkepov {\it et al.} \cite{Esirkepov} at the extremely high intensity of $I>10^{23}\textrm{W}\textrm{cm}^{-2}$. The key principle is using the ultra-strong radiation pressure to accelerate protons to relativistic energy during few laser cycles before the rapid growth of the rear sheath field and the heavy decompression of the foil, so that the protons can quickly catch up with the electrons in an accelerating plasma slab. Recently, Zhuo {\it et al.} \cite{Zhuo}, discussed a dual-peaked electrostatic field acceleration for production of quasimonoenergetic ion beams using LP pulses. In reality, the situation they discuss is qualitatively the same as the laser-piston regime in \cite{Esirkepov}, as they also use a very high laser intensity ($5.5\times10^{22}\textrm{W}\textrm{cm}^{-2}$) resulting in proton acceleration to relativistic energy within 2 laser cycles. There is however a clear interest in identifying regimes where RPA can dominate the acceleration at intensities more reachable with current or near term laser systems employing LP laser pulses. 

In this letter, we identify a new RPA dominated regime of quasimonoenergetic ion beam generation for LP laser pulses at intensities of $10^{20}-10^{21}\textrm{Wcm}^{-2}$. In this regime, with an appropriate coupling condition between foil thickness, laser intensity and pulse duration satisfied, RPA can be dominant during  the competition between RPA and TNSA mechanisms in a hybrid acceleration stage. Such a coupling condition has been analytically derived and verified by two-dimensional (2D) particle-in-cell (PIC) simulations. These show that $1.26 \textrm{GeV}$ quasimonoenergetic $\textrm{C}^{6+}$ ion beam with energy spread $\sim0.1\textrm{GeV}$ and angular divergence $<5^{\circ}$ is obtained by irradiation of $80\textrm{nm}$ ultrathin foils with LP pulses at intensities $10^{21}\textrm{Wcm}^{-2}$. As expected \cite{Bin(2010),Robinson,Yu,Bin(2011)}, the simulations also show that a multispecies foil helps to stabilize the LS RPA component in the hybrid acceleration of lighter ions, resulting in more pronounced spectral peaks, similar to what observed from "pure" RPA using CP lasers. 

When a LP laser pulse irradiates a solid foil target, the electrons in the skin depth of the front surface are accelerated by the ponderomotive force \cite{Paul}  
\begin{eqnarray}
\label{eq:pond}
f_p = -\frac{e^2}{4m_e\omega^2}\nabla |E_L|^2(1-\cos2\omega t),
\end{eqnarray}
which includes non-oscillating and oscillating terms. In Eq. (\ref{eq:pond}), $E_L$ is the amplitude of the laser field oscillating at frequency $\omega$, $m_e$ and $e$ are electron mass and charge, $t$ is the time. Electrons are pushed inward by the non-oscillating ponderomotive term and pile up in a compressed layer, inducing a charge separation electric field $E_z$ which accelerates ions in a pistonlike manner, in the process known as HB RPA \cite{Robinson(2008),Bin(2009),Bin(2010)}.  From the momentum balance, the HB velocity is obtained as \cite{Robinson(2008)}
\begin{eqnarray}
\label{eq:vb}
\frac{v_b}{c}=\frac{\sqrt{n_cm_e/n_im_i}a}{1+\sqrt{n_cm_e/n_im_i}a},
\end{eqnarray} 
where $n_i$ and $m_i$ are ion density and mass, $n_c=m_e\omega^2/e^2$ is the critical density, $a=e|E_L|/mc\omega$ is the laser normalized amplitude and $c$ is light speed. Assuming that the laser field is perfectly reflected, the ions can obtain the maximum velocity $v_{i,front}=2v_b/(1+v_b^2/c^2)$.

However, due to the oscillating ponderomotive term, a fraction of the electrons in the skin depth are heated rapidly by the $\mathbf{j}\times\mathbf{B}$ effect and travel through the target at speed close to $c$. A hot electron cloud forms at a distance of the Debye length $\lambda_h=(k_BT_h/4\pi n_he^2)^{1/2}$ away from the target rear edge, which induces a sheath electric field accelerating ions according to TNSA mechanism. This sheath field scales as the hot-electron temperature $T_h$ multiplying the effective density in the cloud $n_h$,  i.e., $E_{sheath}\approx(4\pi n_hT_h)^{1/2} $. Based on the expanding plasma model, the maximum ion velocity obtained by TNSA can be estimated as \cite{Mora,Fuchs}
\begin{eqnarray}
\label{eq:rear}
v_{i,rear}= C_h[2\ln (\omega_{pi}t)+\ln2-1],
\end{eqnarray}
where $C_h=(T_h/m_i)^{1/2}$ and $\omega_{pi}=(4\pi n_{h}Ze^2/m_i)^{1/2}$ are the hot electron sound speed and the local ion plasma frequency. Furthermore, for a solid foil target of steep density gradient, $T_h$ is obtained as \cite{Chrisman} $T_h=(\gamma n_c/n_{0e})^{1/2}[(1+a^2)^{1/2}-1]m_e c^2$, where $\gamma=(1+a^2)^{1/2}$.

If the foil thickness is less than half of the pulse length, i.e., $l_0<c \tau_L/2$, but thick enough to prevent laser from punching through, it has been shown \cite{Sentoku} that the recirculation of hot electrons through the target increases the effective hot electron density, resulting in enhanced rear sheath field and ion TNSA. The hot-electron recirculation also leads to rapid foil deformation, breaking the equilibrium condition required for LS RPA. Therefore, TNSA usually dominates for ion acceleration by LP laser pulses. The ion acceleration in the sheath field will end achieving the terminal velocity of Eq. (\ref{eq:rear}) after the laser pulse is over. 

\begin{figure}\suppressfloats
\includegraphics[width=9cm]{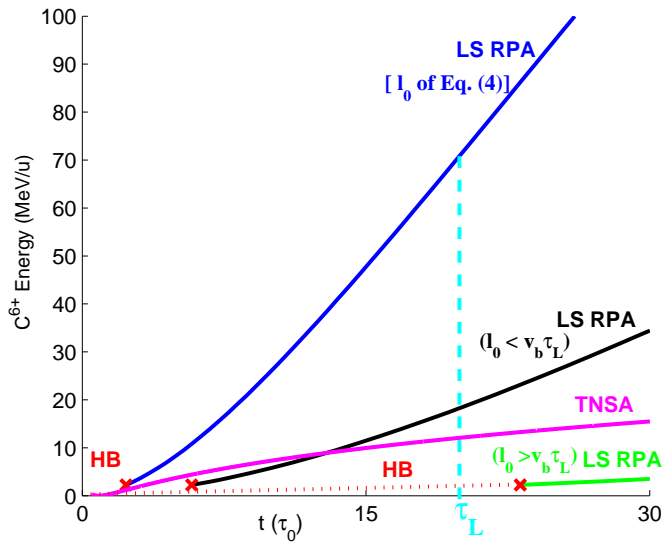}
\caption{\label{fig:analytical} Three regimes for ion acceleration from thin foils by LP laser pulses, based on Eqs. (\ref{eq:vb})-(\ref{eq:LSRPA}): RPA-dominated (blue and purple lines), competing of RPA andTNSA (black and purple), and TNSA-dominated (purple and green), where the cross point corresponds to the transition time from HB RPA to LS RPA.} 
\end{figure}

To achieve RPA dominance with LP laser pulse, a first condition is to use a foil target sufficiently thin that the hole-boring can reach the foil rear surface early enough within the laser pulse duration, that is, $l_0<v_b\tau_L$. Under this condition, the ions from the foil front side accelerated by HB RPA combine together with those undergoing TNSA at the foil rear and experience a hybrid TNSA-LS acceleration. Furthermore, to ensure that RPA dominates over TNSA in the hybrid acceleration, the condition $v_{i,front}>v_{i,rear}$ should be satisfied at the time $t_{hb}=l_0/v_b$ when HB reaches the foil rear surface. This indicates that the ions accelerated by HB RPA can overtake the TNSA-accelerated ions in the LS stage. From Eqs. (\ref{eq:vb}) and (\ref{eq:rear}), we obtain that the coupling condition of the RPA dominance is
\begin{eqnarray}
\label{eq:condition}
\frac{1}{\pi}\frac{n_c}{n_0}a < \frac{l_0}{\lambda}<\frac{1}{2\pi}\!\sqrt{\!\frac{n_c}{\gamma n_0}}a\exp[\sqrt{\!\frac{Zn_c}{n_0}}\!\frac{a}{\sqrt{\gamma\!-\!1}}(\frac{n_0}{\gamma n_c})^{1/4}\!+\!\frac{1\!-\!\ln\!2}{2}]
\end{eqnarray}
where $n_0$ is initial foil electron density and $n_h=\gamma n_c$ is assumed to be the effective hot electron density; $\lambda$ is laser wavelength; $A$ and $Z$ are ion mass and charge number satisfying $n_0=Zn_i$ and $m_i=Am_p$. The left part of the condition (\ref{eq:condition}) satisfies the requirement to avoid complete electron blown-out from the foil target \cite{Macchi(2009),Bin(2010)}.  

If the condition (\ref{eq:condition}) is satisfied, after the HB ends the ion beam acceleration is dominated by RPA in LS stage where TNSA contributes only as a small leakage from the rear surface of the accelerated slab. The peak velocity of ion motion in the LS can be described \cite{Robinson(2008),Bin(2009)} as 
\begin{eqnarray}
\label{eq:LSRPA}
\frac{dv_i}{dt}=\frac{1}{2\pi n_im_il_0}\frac{|E_L|^2}{\gamma^3}\frac{1-v_i/c}{1+v_i/c}
\end{eqnarray}
where $|E_L|\!=\!(m_ec\omega/e)a$. According to Eqs. (\ref{eq:vb})-(\ref{eq:LSRPA}), Fig. \ref{fig:analytical} plots three possible cases of ion hybrid acceleration from foils by LP laser pulses. If an ultrathin foil is used with the condition (\ref{eq:condition}) satisfied, ion acceleration is in the RPA dominated regime (blue line), where high-energy quasimonoenergetic ion beams can be obtained. If the foil is thicker with the condition (\ref{eq:condition}) violated but thin enough to satisfy $l_0<v_b\tau_L$, ion acceleration takes place in a hybrid stage where RPA and TNSA compete (black and purple lines). Since the energy increase of RPA grows more rapidly than TNSA, quasimonoenergetic ion beams may be obtained after a long acceleration time, that is, a long pulse duration $\tau_L$ is required. If the foil is further thicker with $l_0>v_b\tau_L$, ions undergo pure TNSA acceleration (purple line). Therefore, quasimonoenergetic ion beams can be obtained by using either an ultrathin foil target or a long laser pulse even at intensities much lower than those discussed in \cite{Esirkepov,Zhuo}.

\begin{figure}\suppressfloats
\includegraphics[width=14.0cm]{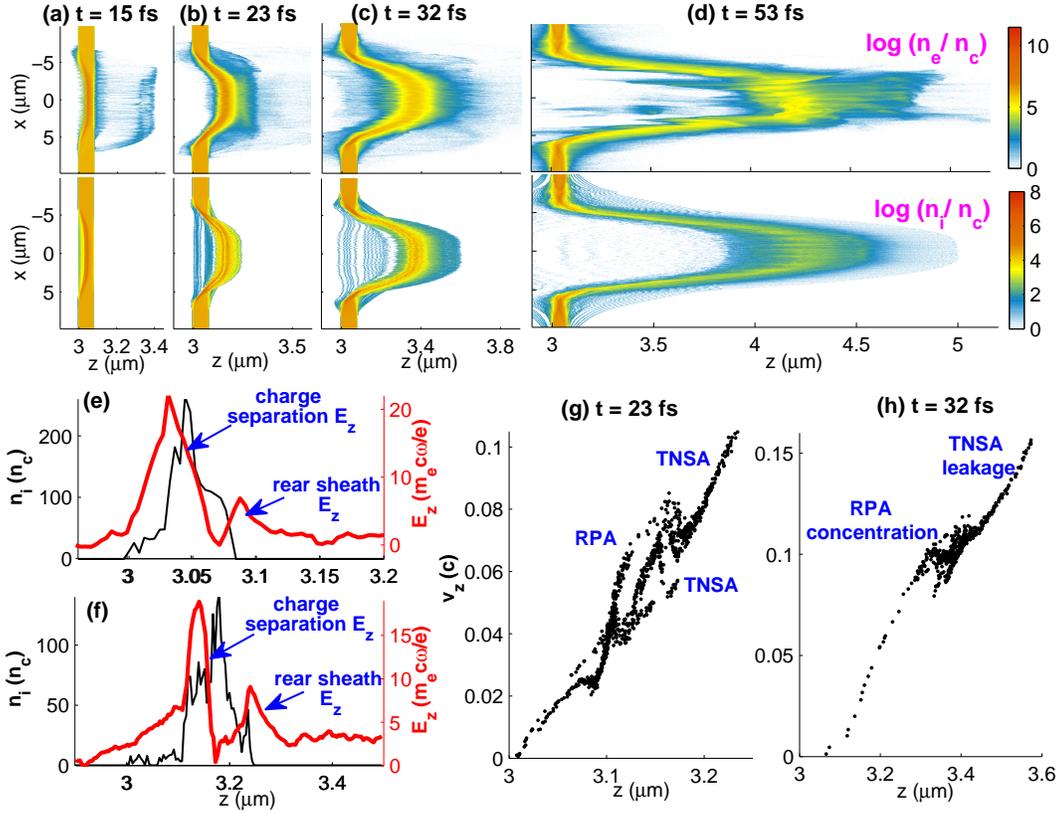}
\caption{\label{fig:map} (a)-(d) Electron ($n_e$) and ion ($\textrm{C}^{6+}$, $n_i$) densities at respectively $t\!=\!15$, $23$, $32$ and $53\textrm{fs}$ for a $80\textrm{nm}$ fully-inoized carbon foil irradiated by LP laser pulses at intensity $I_0\!=\!3\!\times\!10^{21}\textrm{W}/\textrm{cm}^2$, wavelength $\lambda=1.0\mu\textrm{m}$ and duration $\tau_L=20\tau_0=67\textrm{fs}$, where the foil electron density is $600n_c$; (e) and (f) the longitudinal profile of $n_i$ and electrostatic field $E_z$ at $t=15$ and $23\textrm{fs}$, which shows ions are accelerated by a hybrid effect of the charge separation field and the rear sheath field, i.e., undergoing hybrid RPA-TNSA; (g) and (h) $\textrm{C}^{6+}$ phase space distributions at $t=23$ and $32\textrm{fs}$ for particles with $|x|<2\mu\textrm{m}$, which show the typical characteristics of hybrid RPA-TNSA mechanism.} 
\end{figure}

In order to verify the different regimes discussed above, 2D PIC simulations are run with the code "ILLUMINATION". We choose a LP laser pulse with $I_0=3\times10^{21}\textrm{W}\textrm{cm}^{-2}$ ($a\!=\!47$), $\lambda\!=\!1.0\mu\textrm{m}$ and super-Gaussian intensity distribution $\exp[-(r/r_0)^4]$ propagating along z-axis, where $r_0\!=\!5\mu\textrm{m}$ is the spot radius. The laser pulse has a trapezoidal temporal profile of duration $20\tau_0$ ($\tau_0\!=\!2\pi/\omega$), consisting of a plateau of $18\tau_0$ and rising and falling times of $1\tau_0$ each. The fully-ionized carbon foil with electron density $n_0\!=\!600n_c$, $\textrm{C}^{6+}$ density $n_i\!=\!100n_c$ and thickness $l_0\!=\!80\textrm{nm}$ is located at $z\!=\!3\mu\textrm{m}$, where the condition (\ref{eq:condition}) is satisfied. In the simulations, $5000$ cells along z-axis and 3072 cells transversely along x-axis constitute a $20\times24.576\mu\textrm{m}$ simulation box. Each foil cell is filled up with $576$ electrons and $96$ $\textrm{C}^{6+}$ ions.

Figures \ref{fig:map}(a)-\ref{fig:map}(d) show density maps of electrons ($n_e$) and $\textrm{C}^{6+}$ ($n_i$) at time $t=15$, $23$, $32$ and $53\textrm{fs}$. Part (a) corresponds to the initial HB stage. On the one hand, due to the $\mathbf{j}\times\mathbf{B}$ heating, the hot electrons of temperature $T_h=(\gamma n_c/n_{0e})^{1/2}[(1+a^2)^{1/2}-1]m_e c^2\approx7\textrm{MeV}$ quickly travel through the target and form a hot electron sheet of density about $11n_c$  at a distance of their Debye length $\lambda_h\!\sim\!0.2\mu\textrm{m}$ away from the foil rear surface. From Fig. \ref{fig:map}(e) we see that a strong sheath electric field $E_{sheath}\!=\!6m_ec\omega/e\!=\!1.93\times10^{13}\textrm{V}/\textrm{m}$ is induced which accelerates ions from the foil rear side. On the other hand, electrons at the front side are pushed inward by the non-oscillating radiation pressure and pile up in a compressed layer, leading a much more intense charge separation electric field $E_{separation}=22m_ec\omega/e=7.0\times10^{13}\textrm{V}/\textrm{m}$, seen in Fig. \ref{fig:map}(e), which accelerates ions from the foil front side. 

\begin{figure}\suppressfloats
\includegraphics[width=11cm]{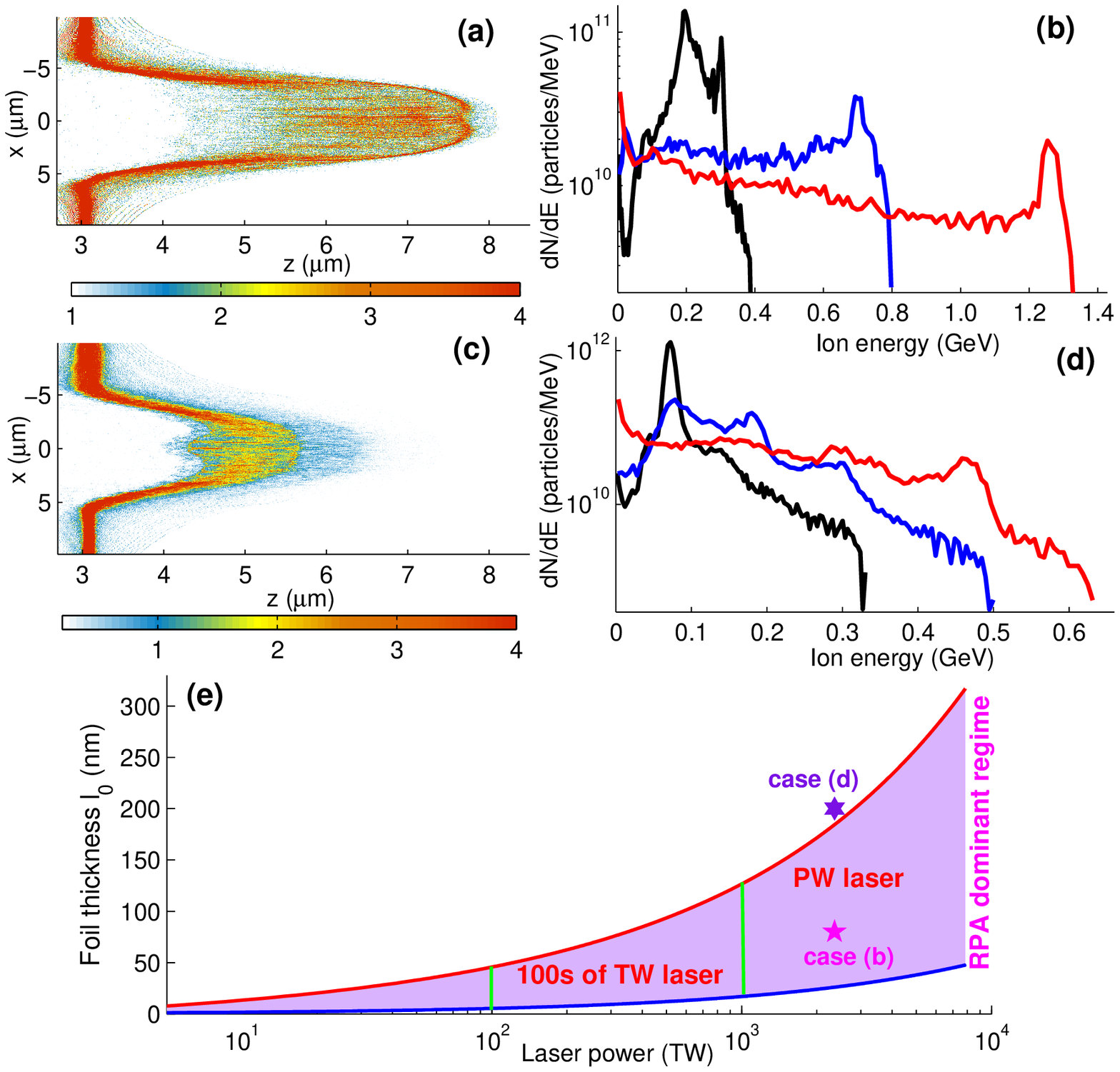}
\caption{\label{fig:spectra} (a) $\textrm{C}^{6+}$ density at $t=83\textrm{fs}$ when the laser pulse ends for the acceleration of Fig. \ref{fig:map}; (b) $\textrm{C}^{6+}$ energy spectra at $t=53$ (black), $68$ (blue) and $83\textrm{fs}$ (red) for Fig. \ref{fig:map}, which have a well-defined peak as RPA is dominant in the hybrid acceleration. (c) and (d) are those for the case by increasing foil thickness to $l_0\!=\!200\textrm{nm}$ where condition (\ref{eq:condition}) is violated, showing no pronounced peak as RPA is not dominant. (e) shows the scaling of the required foil thickness for the RPA dominated regime with the laser power by fixing laser pulse duration at $66\textrm{fs}$ and wavelength $1.0\mu\textrm{m}$, where the target density are the same Fig. \ref{fig:map}. The corresponding points for the spectral cases of (b) and (d) are also respectively marked in (e).}
\end{figure}

Figure \ref{fig:map}(b) corresponds to the transition time when HB reaches the foil rear surface. We see that the compressed electron layer catches up with the rear hot electron sheet; correspondingly the ions that undergo HB RPA also keep up with those at the rear which experience TNSA because $E_{separation}\!>\!E_{sheath}$. At this stage, a majority of ions are bunched together undergoing RPA, but they are preceded by a small number of TNSA ions. This ion bunch is accelerated by a combination of the charge separation field at its front and the sheath field at its rear, see Fig. \ref{fig:map}(f), undergoing a hybrid LS-TNSA acceleration. Since the condition (\ref{eq:condition}) is satisfied, $E_{separation}$ remains always larger than $E_{sheath}$, resulting in quasimonoenergetic ion beams in Fig. \ref{fig:map}(c). This hybrid acceleration dynamics can also be seen in the ion phase plots \ref{fig:map}(g) and \ref{fig:map}(h): a majority of ions undergo typical RPA "rolling-up" and concentration, whereas a small number of ions precede like a leakage due to TNSA.

Figures \ref{fig:spectra}(a) and \ref{fig:spectra}(b) show both a plot of  $\textrm{C}^{6+}$ density at $t\!=\!83\textrm{fs}$ when the laser pulse ends and its energy spectra at $t\!=\!53$, $68$ and $83\textrm{fs}$. We see that a quasimonoenergetic $\textrm{C}^{6+}$ beam with a well defined peak energy of $1.26\textrm{GeV}$, density $4n_c$ and divergence $<5^{\circ}$ is obtained at $t\!=\!83\textrm{fs}$ by the dominant RPA. The low-energy trailing part in the energy spectrum is originated from an extended area (wings) of the foil around the laser focal spot, where the laser field is weaker with the condition (4) violated and the transverse instabilities\cite{Pegoraro(2007)} grows up quickly, so most likely TNSA dominates. Figs. \ref{fig:spectra}(c) and \ref{fig:spectra}(d) show plots for a foil thickness of $200\textrm{nm}$ where condition (\ref{eq:condition}) is violated. We can see that the energy spectrum shows no pronounced peak because RPA is not dominant in the hybrid acceleration. 

In order to emphasize the practical utility of this new regime, we present the scaling of the required foil thickness $l_0$ with the laser power in Fig. \ref{fig:spectra}(e), where the laser pulse duration of $66\textrm{fs}$ and wavelength $1.0\mu\textrm{m}$ are fixed, and the same foil target density as above is chosen. We can see that the RPA dominated regime can be achieved by using ultrathin foils (10s of nm) on the contemporary high-contrast 100s of Terawatt and/or near-term Petawatt laser systems. Our recent experiment \cite{kar(2011)} has also observed narrow band features in both proton and $\textrm{C}^{6+}$ spectra from 100nm copper foil by Vulcan laser (energy 200J, duration $\sim500\textrm{fs}$) at Rutherford Appleton Laboratory, which has further verified the regime. Furthermore, note that the trapezoidal temporal profile of laser pulse chosen in the simulations are only intended to to show the exact validity of the condition (4) as the latter is derived by assuming $a$ is a constant. The regime itself in principle is not sensitive to laser pulse profile, which only requires the general condition $v_{i,front}>v_{i,rear}$ [from Eqs. (2) and (3)] to be satisfied. 

\begin{figure}\suppressfloats
\includegraphics[width=12cm]{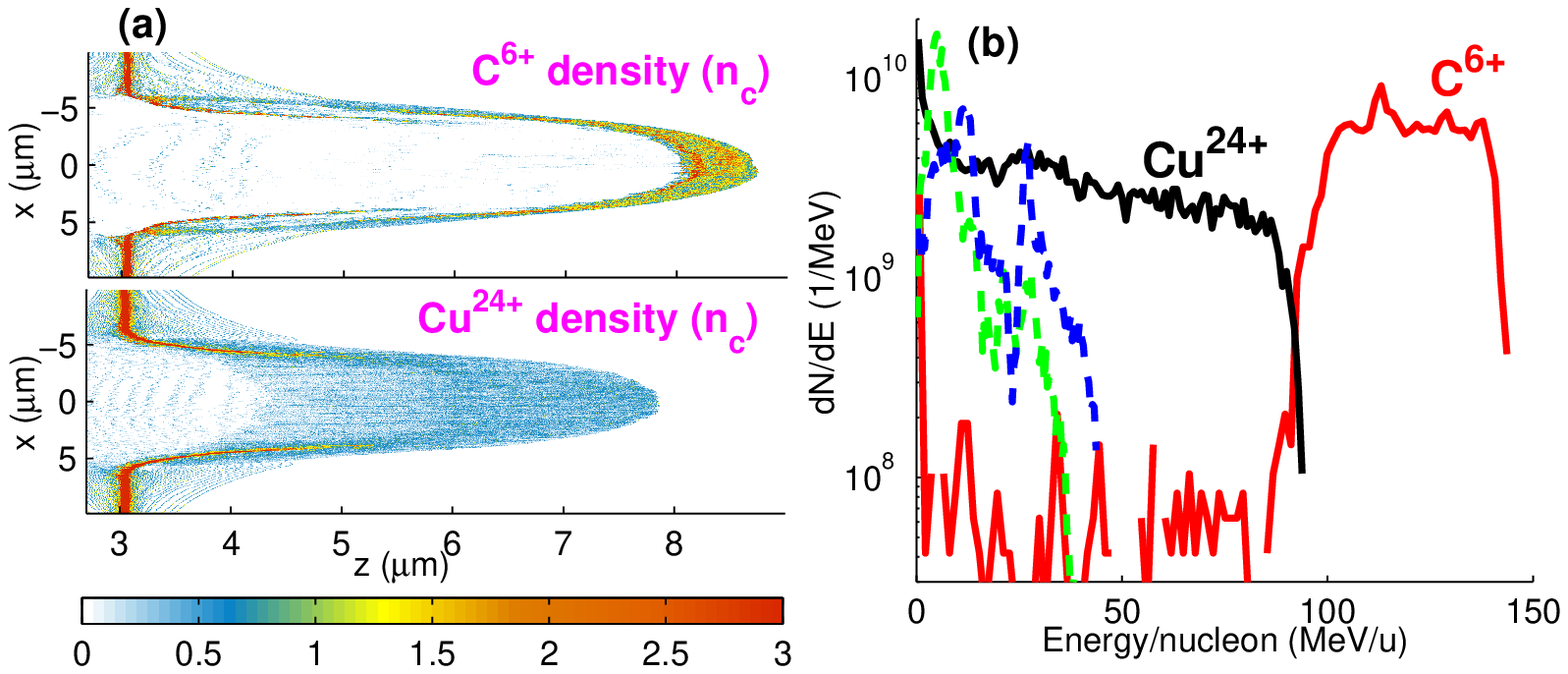}
\caption{\label{fig:multispecies} Ion acceleration from a $80\textrm{nm}$ fully-ionized copper-carbon mixed foil target by LP laser pulse, where the density ratio between $\textrm{Cu}^{24+}$ and $\textrm{C}^{6+}$ is $5:1$. The total mass density and other parameters are the same as those in Fig. \ref{fig:map}. (a) is the plot of both $\textrm{Cu}^{24+}$ and $\textrm{C}^{6+}$ densities at $t\!=\!83\textrm{fs}$ when the laser pulse ends; (b) shows the corresponding energy spectra. The dashed blue and green lines are the energy spectra for $\textrm{C}^{6+}$ in the cases of thicker foil with $l_0=200$ and $800\textrm{nm}$. }
\end{figure}

A typical experimental target is always a solid foil contaminated with lighter ions (carbon and hydrogen).  As expected \cite{Bin(2010),Robinson,Yu,Bin(2011)}, a multispecies target helps to stabilize LS RPA of the light ion species so that a more pronounced peak in the energy spectrum can be achieved. Fig. \ref{fig:multispecies} shows density map and energy spectrum of the acceleration case by choosing a fully-ionized copper foil with carbon contaminants, where the density ratio between $\textrm{Cu}^{24+}$ and $\textrm{C}^{6+}$ species is $5:1$. Their total mass densities and the other parameters are the same as above. We see that a more pronounced monoenergetic $\textrm{C}^{6+}$ beam is obtained with density $3n_c$, divergence $<5^{\circ}$ and higher peak energy $1.44\textrm{GeV}$ ($120\textrm{MeV}/\textrm{u}$), whereas the $\textrm{Cu}^{24+}$ species has insufficient co-moving electrons and undergoes Coulomb explosion resulting in a broadened energy spectrum. Note that here the particle number ($dN/dE$) is about $10^{10}\textrm{MeV}^{-1}$, much larger than peaks obtained from the improved TNSA \cite{Nature(2006)}.  

To summarize, we have reported a new RPA dominated regime of quasimonoenergetic ion beam generation from ultrathin foils for LP laser pulses at intensities two-orders of magnitude than earlier predicted \cite{Esirkepov}. A criterion for this regime has been identified analytically and verified by 2D PIC simulations, which indicates that the RPA dominated regime can be achieved with LP laser pulses by using either a sufficiently thin foil target or a long enough laser pulse. It is shown by 2D PIC simulations that quasimonoenergetic $\textrm{C}^{6+}$ ion beams with peak energy about $1.26\textrm{GeV}$, particle number $10^{11}$ and beam divergence $<5^{\circ}$ are obtained from $80\textrm{nm}$ carbon foils by LP laser pulses at intensities of $10^{21}\textrm{Wcm}^{-2}$.

We acknowledge helpful discussions with A. Macchi, J. Schreiber and B. Dromey.

\newpage

\end{document}